\documentclass[conference]{IEEEtran}
\usepackage{cite}
\usepackage{amsmath,amssymb,amsfonts}
\usepackage{algorithmic}
\usepackage{graphicx}
\usepackage{textcomp}
\usepackage{xcolor}
\usepackage{fancyhdr}
\usepackage[hyphens]{url}
\usepackage{soul}

\def\BibTeX{{\rm B\kern-.05em{\sc i\kern-.025em b}\kern-.08em
    T\kern-.1667em\lower.7ex\hbox{E}\kern-.125emX}}

\newcommand{\iscasubmissionnumber}{504}
\renewcommand{\tilde}[0]{$\sim$}

\newcommand{\hide}[1]{ }
\newcommand{\authorhide}[1]{ }
\newcommand{\hints}[1]{ }
\newcommand{\ee}[0]{$^\dagger$}
\newcommand{\cs}[0]{$^\S$}

\fancypagestyle{firstpage}{
  \fancyhf{}

\hide{
  \fancyhead[C]{\normalsize{ISCA 2020 Submission
      \textbf{\#\iscasubmissionnumber} \\ Confidential Draft: DO NOT DISTRIBUTE}}} 
  \fancyfoot[C]{\thepage}
}

\title{Semantic prefetching using forecast slices} 

\author{Leeor Peled\ee ~~~~~~~~ Uri Weiser\ee ~~~~~~~~ Yoav Etsion\ee\cs\\
	\ee{}Electrical Engineering ~~~~~~~~~ \cs{}Computer Science \\
	Technion --- Israel Institute of Technology\\
	\normalsize\{leeor@tx, uri.weiser@ee, yetsion@tce\}.technion.ac.il}

\begin{document}
\maketitle
\thispagestyle{firstpage}
\pagestyle{plain}


\begin{abstract}
	

Modern prefetchers identify memory access patterns in order to predict future accesses. 
However, many applications exhibit irregular access patterns that do not manifest any form of spatio-temporal locality in the memory address space. Such applications usually do not fall under the scope of existing prefetching techniques,
which observe only the stream of addresses dispatched by the memory unit but not the code flows that produce them. 
Similarly, temporal correlation prefetchers detect recurring relations between accesses, but do not track the chain of causality in program code that manifested the memory locality. 
Conversely, techniques that are code-aware, like runahead execution, are limited to the basic program functionality and are bounded by the machine depth. 


In this paper we show that contextual analysis of the code flows that generate memory accesses can detect recurring code patterns and expose their underlying semantics even for irregular access patterns. Moreover, program locality artifacts can be used to enhance the memory traversal code and predict future accesses. 
We present the \textit{semantic prefetcher} that 
analyzes programs at run-time and learns their memory dependency chains and address calculation flows. 
The prefetcher then constructs \textit{forecast slices} 
and injects them at key points to trigger timely prefetching of future contextually-related iterations. 
We show how this approach takes the best of both worlds, augmenting code injection with forecast functionality and relying on context-based temporal correlation of code slices. 
This combination allows us to overcome critical memory latencies in access patterns that are currently not covered by any other prefetcher.

Our evaluation of the semantic prefetcher using an industrial-grade, cycle-accurate x86 simulator shows that the semantic prefetcher improves performance by 24\% on average over SPEC 2006 (with outliers up to 3.7$\times$), and 16\% on average over SPEC 2017 (with outliers up to 1.85$\times$), using only \tilde6KB of structures. 
	
\end{abstract}

\section{Introduction}

Existing prefetchers are designed to analyze the memory access stream and identify specific types of access patterns, ranging from sequential and strided ones to traversals over linked data structures. 
Most of them target \emph{spatio-temporal locality} and \emph{temporal correlation} between addresses or address-space artifacts (e.g., address deltas), based on the observation that temporally or spatially adjacent accesses tend to repeat~\cite{PrimerHWPref}. 

Many applications, however, make use of data structures and algorithms whose physical layout and data access patterns are not plainly observable in the memory address-space domain (e.g., linked lists, arrays of pointers, sparse graphs, cross-indexed tables) and require deeper analysis in order to understand the causal relations 
between accesses to objects in memory. These causal relations 
may involve complicated arithmetic computations or a chain of memory dereferences.  
Such relations between accesses exhibit \textit{semantic locality}~\cite{context_pref}
if they represent consequential steps along a data structure or an algorithm. These steps are characterized by the existence of some program code flow that traverses from one data object to the next 
\footnote{Spatio-temporal locality represents a specific case where the relations are purely arithmetic and can be detected through address comparison, but semantic locality encompasses all forms of algorithmic and data structural relations (such as proximity within linked data structures or connectivity in cross-indexed tables).}.
The set of all semantic relations within a program can be said to \textit{span} its data structures, describing all the steps that the program may employ to walk through them. 

In this paper we argue that the semantic relations between memory accesses can be represented through the code segments (referred to as code slices) that generate the memory traversals.
The set of all slices effectively forms an abstract guide to the program's data layout, but we can further combine or extrapolate these flows to create \textit{forecast slices} with more complex ``lookahead'' semantics that can predict program behavior over longer periods.

Following our observation, we present the \textit{semantic prefetcher} that dynamically constructs and injects prefetching code for arbitrary memory traversals. The prefetcher analyzes program code at run-time, identifies the dependency chains forming all address calculations, and detects locality artifacts within that code based on contextual similarities. The prefetcher then generates compact and optimized forecast slices which are 
code constructs that did not exist in the original program 
and enhance the code to generate longer memory traversal steps 
capable of reaching future iterations.
The semantic prefetcher generates the forecast slices using hardware-managed binary optimization. The slices are constructed to have no lingering side effects. 
Once the prefetcher reaches sufficient confidence in their correctness and structural stability, it injects them at certain interception points to trigger prefetches. 

The semantic prefetcher is fundamentally different from previous prefetchers that aim to reconstruct address relations or code sequences such as temporal-correlation prefetchers~\cite{GHB, SMS2, STMS, Domino} and runahead-based prefetchers~\cite{scouting, slice_proc, precompute, precompute2, runahead_ooo, continuous_runahead, runahead_smt}. 
Unlike temporal correlation prefetchers, which detect correlations between addresses, the semantic prefetcher correlates program states (specific code locations with specific history and context) with the generated code slices. 
Similarly, unlike runahead-based prefetchers that run the program (or its address generation code) in parallel to reach future iterations earlier 
(but are ultimately constrained by finite out-of-order depths), the semantic prefetcher can peek into future iso-context iterations without having to execute everything in the middle.

The semantic prefetcher was implemented on an industrial-grate, cycle-accurate x86 simulator that represents a modern micro-architecture. It provides a 24\% IPC speedup on average over SPEC 2006 (outliers of up to 3.7$\times$), and 16\% on average over SPEC 2017 (outliers of up to 85\%). 

Our contributions in this paper are as follows:
\begin{itemize}

\item
  We present a novel scheme of prefetching using forecast slices. We utilize internal locality artifacts 
  to extrapolate the code slices and create new functional behavior with lookahead semantics. 

\item
 We present the design of the semantic prefetcher that injects forecast slices directly into the execution stream. We describe its architecture: flaky load detection, slice generation, binary optimization, and dynamic prefetching depth control.

\item
 We demonstrate how the forecast slices can reproduce complex patterns prevalent in common applications, and show that these patterns are not addressed by existing prefetchers. 
 
\item 
 We model the semantic prefetcher using a cycle accurate simulator. 
 We show that it outperforms five competing state-of-the-art prefetchers, some of which target irregular access patterns. 
\end{itemize}

The remainder of this paper is organized as follows: Section~\ref{sec:semantic} discusses semantic locality and its manifestation in forecast slices. Section~\ref{sec:arch} presents the semantic prefetcher and its architecture. Section~\ref{sec:methodology} explains the experimental methodology. Section~\ref{sec:evaluation} shows the evaluation results and discussion. Section~\ref{sec:related} describes related work. We conclude in Section~\ref{sec:conclusions}.


\section{Extracting Semantic Locality from memory access patterns}
\label{sec:semantic}
Existing memory prefetchers scan the stream of memory accesses 
and extract spatio-temporal correlations in order to identify patterns and predict future memory accesses~\cite{PrimerHWPref}. Some prefetchres~\cite{GHB, Baer} also associate memory accesses with program context (e.g., instruction pointer) to further refine their predictions.

However, basing predictions solely on the stream of memory accesses that the memory unit emits makes prefetchers oblivious to the underlying program code semantics. 
Indeed, most existing prefetchers ignore the data and control flows that generate the memory access sequences they are meant to detect. 
A small number of exceptional prefetchers 
capable of detecting more elaborate or irregular relations focus only on specific access patterns such as indirect accesses (for e.g., A[B[i]])\cite{IMP} and linked data structures~\cite{Roth98,Roth99,Bekerman99}.


In this section we argue that a more fundamental form of locality can be extracted even when no spatio-temporal locality is present. ~\textit{Semantic locality}~\cite{context_pref, nnpref} correlates memory accesses through their dependency within the program's abstract data layout and usage flow, such as being adjacent steps on a data structure traversal path or being consequential steps in the execution of an algorithm. These accesses do not necessarily exhibit any spatio-temporal correlation. 
While prior work attempted to approximate semantic locality through memoization and correlative program context cues, 
we show that extracting this form of locality requires following the set of operations that constitutes the relation between two memory addresses.
To this end, we define a \textit{code slice} as the minimal subset of the dynamic code preceding a certain memory operation that is required to generate its memory address. Notably, this subset can be described through the data dependency chain that starts with the address calculation, and goes backwards through all relevant sources at each step. 

As semantic locality usually describes program constructs such as data structures or algorithms, the relations it captures often have strong recurrence. Extracting that form of locality can therefore be achieved by analysis of the address-generation code slice between two recurring \textit{consequential} loads. 

The remainder of this section demonstrates how program introspection can generate short, explicit code slices that can be replayed to generate memory accesses, or manipulated to create forecast slices that generate future accesses at an arbitrary distance. These code slices can be injected into the code sequence at run time to issue memory accesses ahead of time. Finally, the memory access stream of typical programs is shown to be adequately covered by a small number of distinct code slices.



\begin{figure}[t]
	\centering
	\includegraphics[width=1.05\columnwidth]{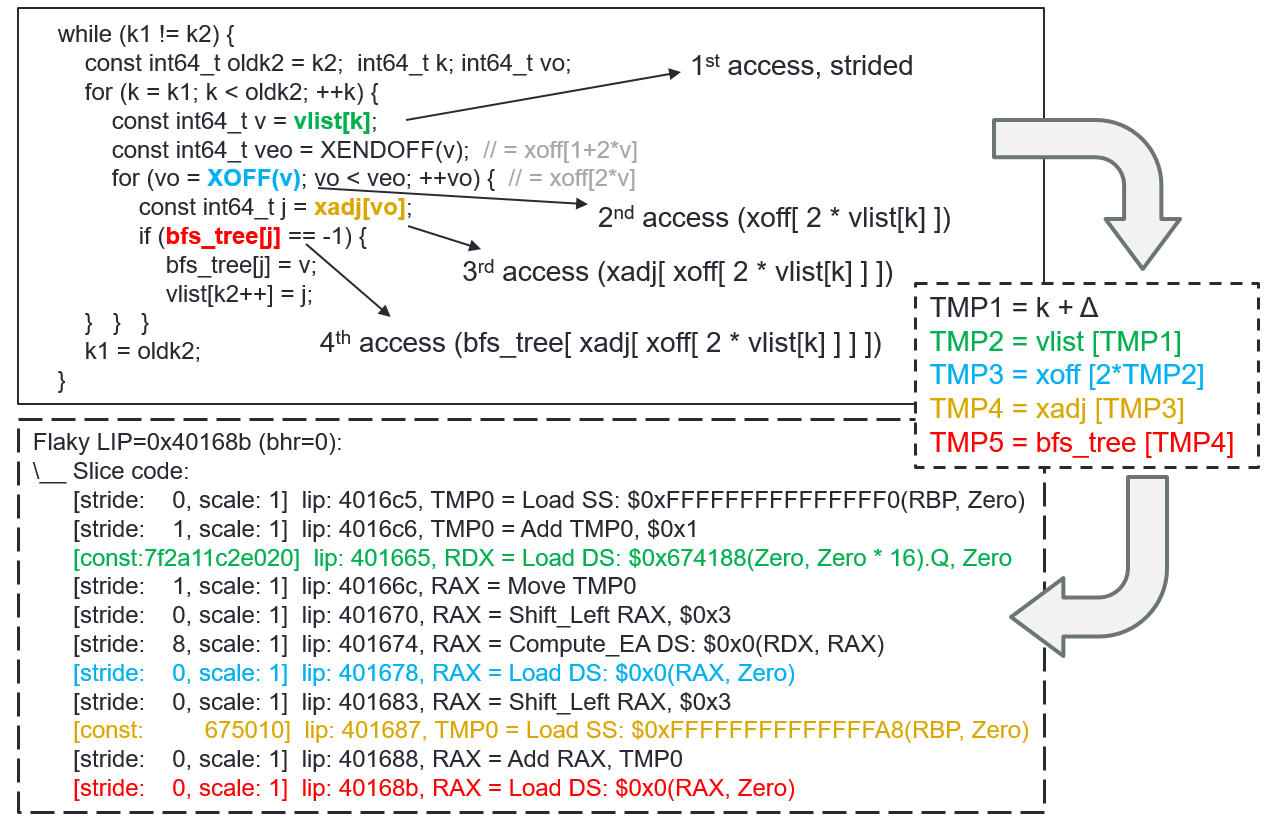}
	\caption{Critical BFS loop in graph500 
	showing a 4-level indirection. The top box shows the source code, the middle shows the pseudo code subset that comprises the slice. The bottom box shows the actual slice generated at run-time.}
	\label{fig:graph500}
\end{figure}

\subsection{How code slices describe dependency chains}
\begin{figure}[t]
	\centering
	\includegraphics[width=0.8\columnwidth]{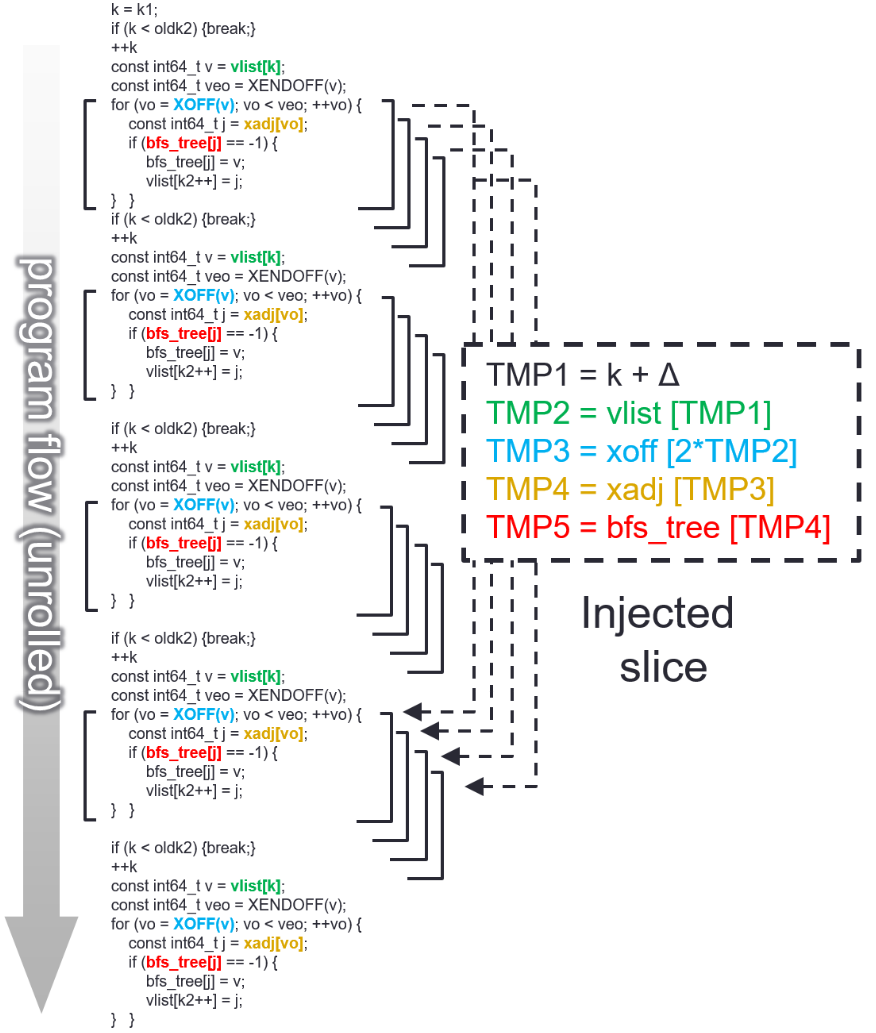}
	\caption{Dynamic flow of graph500 (Figure~\ref{fig:graph500}) unrolled to show a possible lookahead prefetch based on the forecast slice. Changing the stride at the beginning of the slice can predict accesses on far iterations. Furthermore, iterations with similar contexts (such as branch history) will invoke their future counterparts.}
	\label{fig:graph500_flat}
\end{figure}


Memory access patterns can often be tightly incorporated in a way that makes it difficult for a simple address scan to distinguish between them without understanding program semantics.
Figure~\ref{fig:graph500} demonstrates this over a breadth-first search (BFS) code taken from the Graph500 benchmark. The main \textit{while} loop traverses an array of vertices. An internal loop then scans each vertex's outgoing edges to find its neighboring vertices and check their BFS depth.

Notably, the top level access pattern (array ``vlist'') is sequential, but the deeper levels are accessed through data dependent patterns (the edge loop is also sequential but relatively short, making the first edge of each vertex the critical element). These accesses have no spatial locality and very little temporal reuse. Even contextual cues such as the program counter do not help in correlating the accesses.
However, the figure shows that the dependency chain within each iteration, whose accesses are increasingly critical to program performance, can be represented using a short code slice. 

The use of the extracted code slice is demonstrated in Figure~\ref{fig:graph500_flat}. Thanks to the sequential nature of the top loop that exposes spatial locality within the first load in the slice, a simple change in the stride delta can create a forecast slice that predicts accesses in the next iterations at the top loop. 
Overall, the example detailed in Figures~\ref{fig:graph500} and~\ref{fig:graph500_flat} demonstrates how code slices can represent the dependency chain of irregular data structures, and how these slices can generate lookahead semantics within the algorithm.




\subsection{Forecast slice creation}
Tracking all dependency chains for a given load would construct a graph of instructions that may span back to the beginning of the program. To generate concise and useful slices, history tracking is limited by breaking the dependency chain in the following cases:
\begin{itemize}
	\item \emph{Constant values} remaining 
	 static during analysis. 
	\item \emph{Strided values} that were shown to have a constant stride or are produced by a simple add/sub operation with constant or immediate sources.
	\item When a loop wraps around to the same operation where the analysis started from, the dependency chain can usually stop as it would also repeat itself. Linked data structures may iterate a few time to create a deeper chain. 
\end{itemize}

Before the code slice can be used to produce future accesses, it needs to be clean of any side effects. The code is sanitized by performing two final steps: first the destination registers are replaced with temporary ones (which are guaranteed not to be used by the original code) and their occurrences as sources within the slice are renamed accordingly. Second, all memory writes are eliminated from the slice. Since the code was generated through a dependency chain, all writes to memory were added to resolve younger loads reading from the same address. Therefore, a simple memory-renaming may be performed to replace such store-load operations with move operations to a reserved register. For the sake of simplicity partial address overlaps are ignored.


When the base slices are ready, they may be converted into forecast slices. 
To this end, any detected stride is extended by the \textit{lookahead} factor: If a certain operation in the slice was detected to induce a stride of $N$, that stride is replaced by $L\times N$ where $L$ is the current lookahead applied for that slice. This lookahead variable is initialized to point a few iterations ahead, but its value dynamically changes to allow further lookahead based on the average hit depth for that slice. The hit depth is updated dynamically as explained in Section~\ref{sec:arch}. 

\subsection{Data-structure spanning code slices}

\begin{figure}[t!]
	\centering
	\includegraphics[width=0.85\columnwidth]{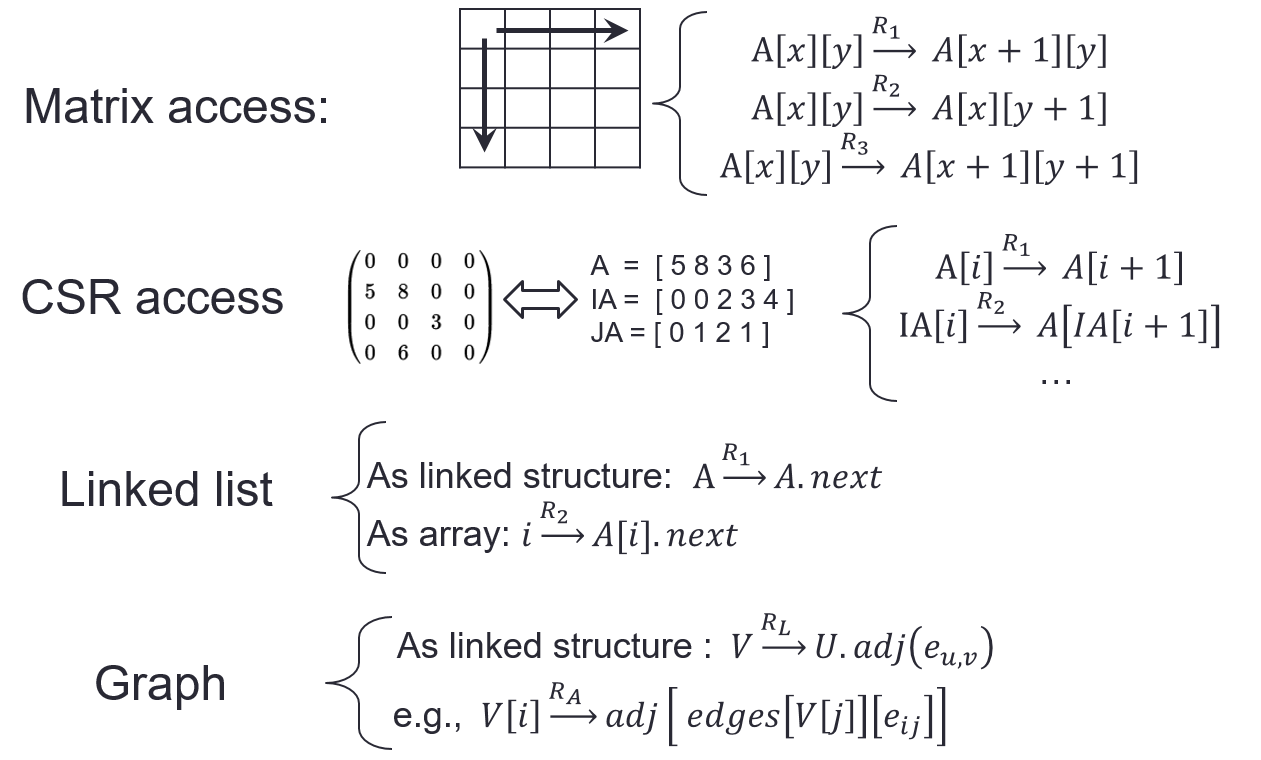}
	\caption{Example data structures and some of their spanning slices in pseudo arithmetic.}
	\label{fig:data_structs}
	\vspace{-1mm}
\end{figure}

Code slice generation is flexible and generalized enough to cover whole applications 
efficiently, with a relatively low amount of code slices.
Any given data structure has a set of operations that define all forms of traversals across it within a given program. We define this set of operations as \textit{spanning} the data structure. Some examples are shown in Figure~\ref{fig:data_structs}.
\hide{
	Figure~\ref{fig:data_structs} demonstrates several common data structures (array of indices, linked list, and some common forms of graphs) and shows for each of them a few examples of pseudo-code functions that generate consecutive memory addresses, spanning these structures.
}

A linked list, for example, is spanned by the action of dereferencing its next elements pointer. A tree is spanned by the actions of descending from a node to any given child. 
The semantic relation must capture all data structures required to complete any recurring traversal step.

\begin{figure}[t]
	\centering
	\includegraphics[width=1.03\columnwidth]{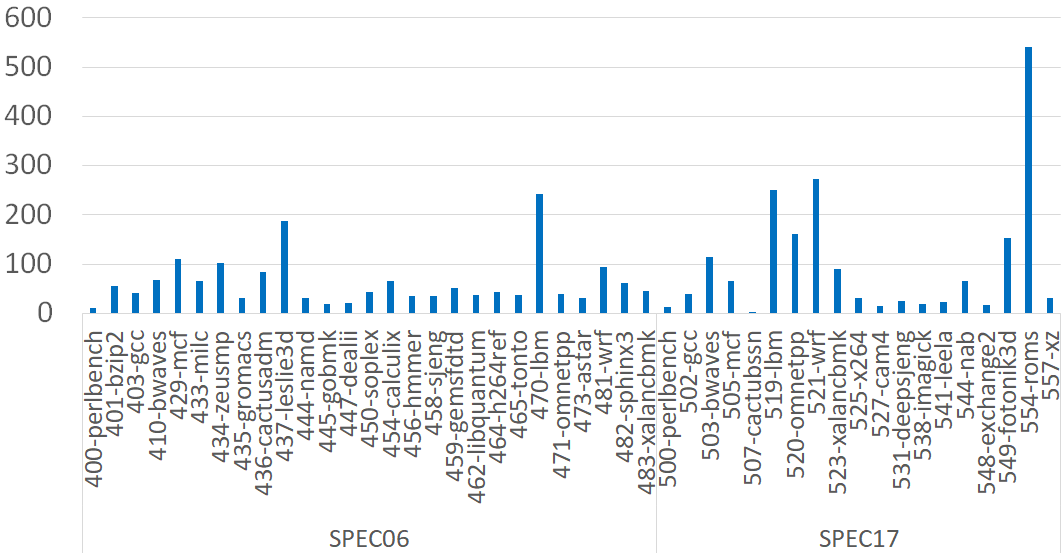}
	\caption{Number of unique slices sampled in SPEC 2006/2017. Collected over slices with at least 1k prefetches sent. Since any recurring load would attempt constructing a slice, this represents the number of slices required for coverage of all recurring loads.}
	\label{fig:slice_count}
\end{figure}

\begin{figure}[t!]
	\centering
	\includegraphics[width=0.95\columnwidth]{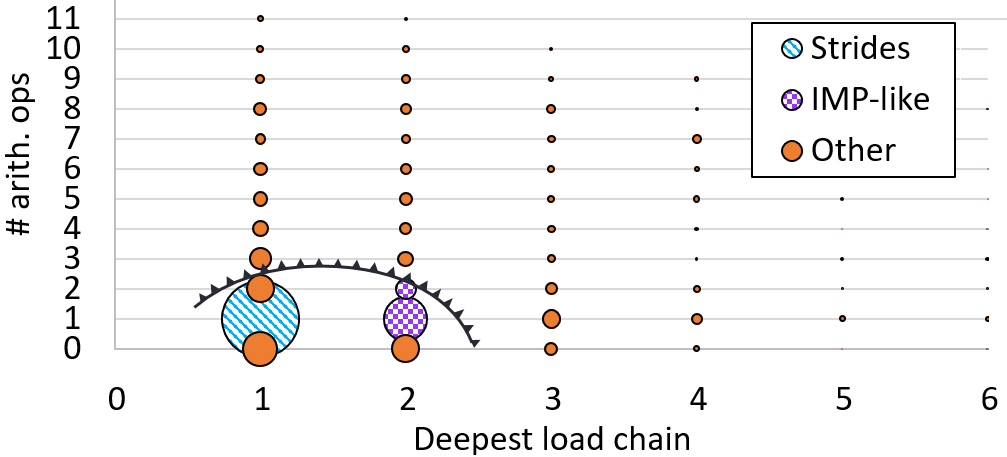}
	\caption{Breakdown of slices in Figure~\ref{fig:slice_count} according to their load-chain depth and number of arithmetic operations, measured over SPEC 2006/2017. The front line marks the rough range covered by existing prefetchers.}
	\label{fig:slice_metrics}
\end{figure}


We demonstrate the effectiveness of code slices in Figure~\ref{fig:slice_count}, which shows the number of unique slices needed to cover all accesses to the main data structures in the 
SPEC 2006 and 2017 benchmarks (sampling methodology is explained in Section~\ref{sec:methodology}). The results were obtained by running the construction flow on each load that has a sufficient level of recurrence (recurring at least three times and passing seven validation phases to confirm that its slice is invariant). We filtered out low-usage strides (below 1k of actual hits on a generated slice). 
Figure~\ref{fig:slice_count} demonstrates the efficiency of memory access coverage of code slices. Specifically, 39 of the 46 benchmarks require only up to \tilde 100 slices to cover all recurring loads, and only one benchmark requires more than 300 slices. This indicates that a prefetcher constructing code slices can cover a large code base with reasonable storage requirements.
The average slice size is 6.6 operations, and the median is 3.5 operations.

Detecting semantic locality through code slices generalizes the existing paradigms of data locality. Figure~\ref{fig:slice_metrics} classifies the code slices (observed in Figure~\ref{fig:slice_count}) according to their memory dereference depth (longest dependent load chain within the slice) and the number of arithmetic operations. The circle sizes indicate relative number of slices within each bucket.
The special case of (1,1) represents slices that have a single arithmetic operation and a single load based on it, which for the most part conform with the common stride pattern. Another interesting case is the (2,1) and (2,2) data points (two loads and one or two arithmetic operation), which includes most examples of array-of-arrays/pointers (A[B[i]] accesses): one index stride, one internal array reference, possible index/pointer math and outer array access. These are potentially covered by IMP~\cite{IMP} or similar dereference based prefetchers like Jump-pointers~\cite{Roth99}. 

Notably, while the largest data points pertain to loads that are addressed by existing prefetchers (37\% of all loads in the figure fall under the stride pattern; 13\% are within the two simple single-dereference patterns), there are still many cases left outside that are not targeted by any existing prefetcher. 
Semantic analysis can cover all these cases using the same mechanism, thereby generalizing existing prefetchers without having to manually design for each use-case. 

\section{Prefetcher architecture}
\label{sec:arch}

\begin{figure}[t!]
	\centering
	\fbox{\includegraphics[width=0.95\columnwidth]{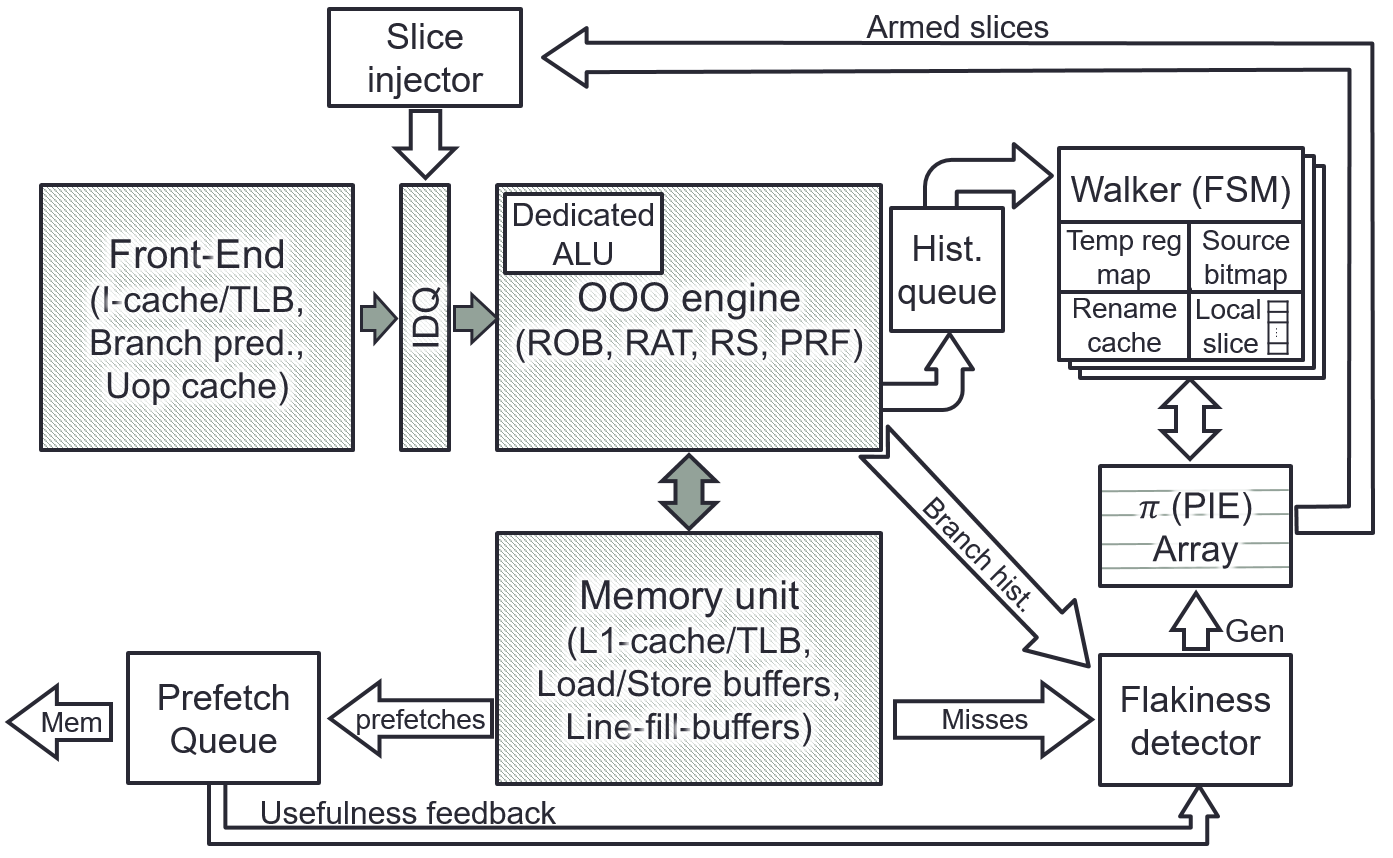}}
	\caption{Semantic prefetcher block diagram (existing core blocks in gray).}
	\label{fig:block_diagram}
	\vspace{3mm}
	\centering
	\fbox{\includegraphics[width=0.95\columnwidth]{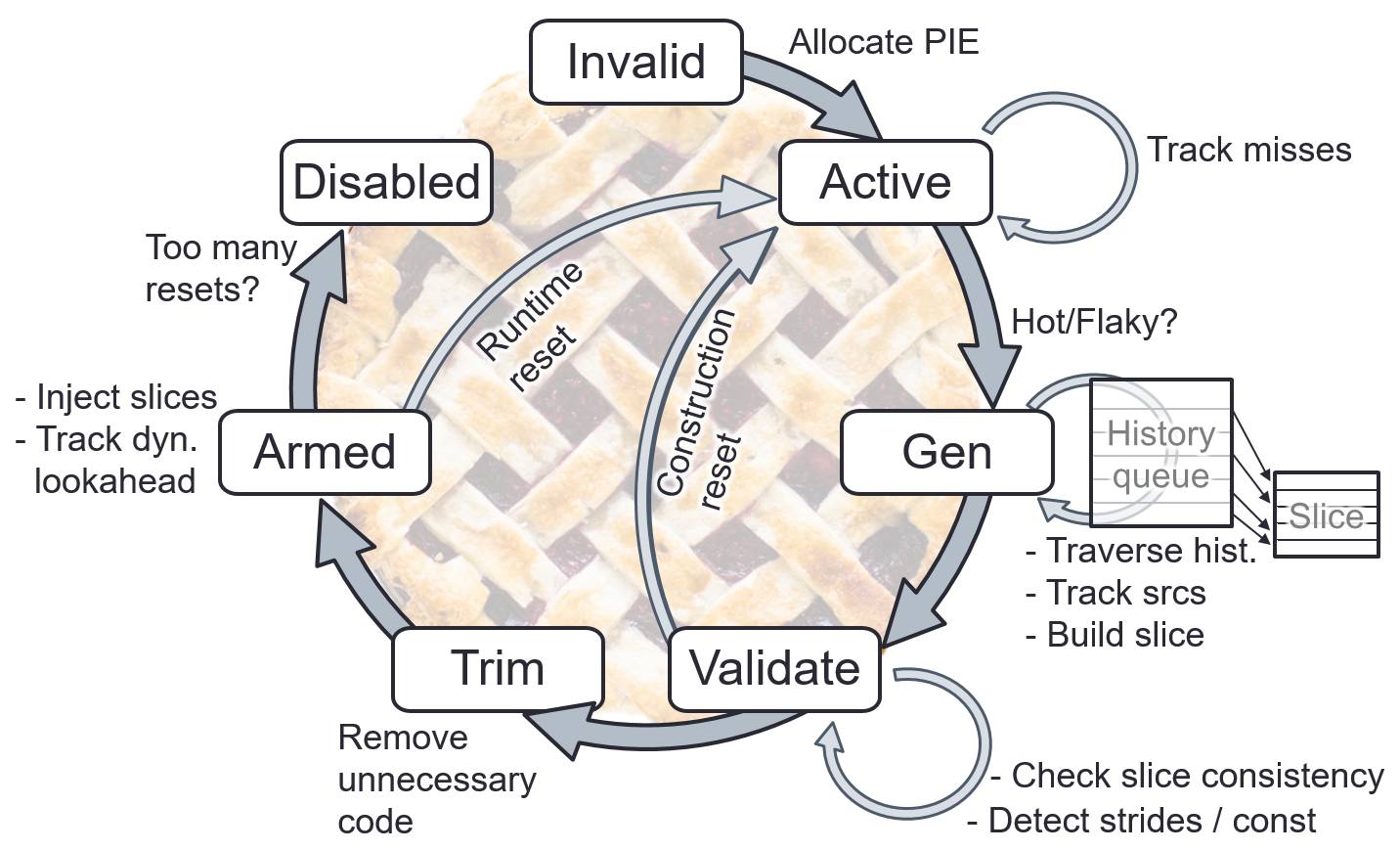}}
	\caption{PIE lifetime flow chart.}
	\label{fig:slice_gen}
	\vspace{-2mm}
\end{figure}

In this section we describe the architecture of the semantic prefetcher. Figure~\ref{fig:block_diagram} shows the high level block diagram of the components:
1) The \textit{flakiness detector} for tracking recurring loads;
2) A cyclic \textit{History queue}, tracking retired code flow;
3) The \textit{prefetch injection entries (PIE) array}, storing slices; 
4) Several \textit{walker} FSMs, generating and validating slices;
5) The \textit{slice injector}.
and 6) The prefetch queue for tracking usefulness and providing feedbacks.


\subsection{Flaky load detection}
The first component is the \textit{flakiness detector}, which is responsible for isolating loads that have both high recurrence and miss rates. The unit identifies and tracks load context by a combination of its instruction pointer (IP) and a branch history register (BHR). 
Our BHR tracks up to 6 recent branches. Each branch is represented by the lowest 4 bits of its IP, with the least significant bit XOR-ed with the binary outcome of that branch (taken or not). Together, the load IP and the BHR represent an instance of load within a specific program context. This method can distinguish between different occurrences within nested loops or complex control flows, which may affect the access pattern and the generated slice. The IP and BHR are concatenated and hashed to create an index that would identify the load throughout the prefetcher mechanisms.

For each load missing the L1, the prefetcher allocates a \textit{prefetch injection entry} (PIE) in the PIE array. These entries serve to track potential loads (within some context) and, if considered useful, construct a slice for them and store it for prefetching. Figure~\ref{fig:slice_gen} describes the life cycle of a single PIE.
Once allocated, the entry starts at the "Active" state. The flakiness detector tracks recurrence and miss rate for each of the active loads. 
Once a PIE has been qualified as flaky (above-threshold miss rate) and hot (high recurrence over a time window) its state switches to "Gen" and it is assigned a walker to construct its slice of code (a PIE slice).

\subsection{PIE slice generation}
The slice that generates the load address consists of a subset of the code path prior to that load. The prefetcher tracks the program code flow at retirement using a cyclic history queue, although in modern processors this can be replaced with existing debug features such as Intel's Real-Time Instruction Tracing (RTIT)~\cite{RTIT}. Once a PIE is switched to "Gen" state and needs to construct a slice it is assigned one of the free walker finite-state-machines (FSMs). The walker traverses the history queue from the youngest instruction (the flaky load itself) to the oldest and constructs the PIE slice.

To track data dependency, the walker uses 1) a \textit{source bitmap} which assigns one bit per register to track the active sources (only general-purpose and flags registers); 2) a \textit{renaming cache} that tracks memory operations for potential memory renaming~\cite{MRN}; 3) a \textit{Temporary register map} that tracks architectural registers replaced with temporary ones. 
The walker also has storage for 16 operations that serves as the local copy of the slice during construction. Finally, the walker has an index pointing to the history queue (for the traversal), an index for the local slice (for construction), and a counter of temporary registers used for memory renaming.

The walker first sets the bits representing the load sources, and then traverses the history queue backwards (from youngest to oldest instruction). 
On each instruction that writes to a register in the bitmap, the walker does the following:
\begin{itemize}
	\itemsep0em 
	\item Pushes the instruction to its local code slice (using an index that starts from the last entry and going backwards from the end of the slice).
	\item Clears the destination register from the sources bitmap marking that its producer has been added.
	\item Sets all the registers corresponding with the current instruction sources. This ensures older operations producing these sources will also be added.
	\item Records the destination value. This will be checked for constants or strides during the next phases.
\end{itemize} 
Loads that are added to the slice record their address and their index within the local slice in the rename cache. This structure can host 16 addresses in the form of a set-associative cache. The walker then performs memory renaming whenever an older store is observed (further along the walk) that matches an adderess in the rename cache. The renaming is done by extracting the index of the matching load from the structure and replacing both the store and the load operations in the slice with a move to and from (respectively) an available temporary register. 
It should be noted that reducing the store/load pair further by moving the store data directly to the load destination is not possible, since the load destination register may be reused between the store and the load (and therefore override the data). 

The walker completes the traversal upon 1) reaching the tail of the cyclic history queue; 2) when there are no longer valid sources marked in the source bitmap; or 3) when the loop completes a round-trip and the same load within the same BHR context is encountered. 
Upon successful completion, the walker switches the PIE to "Validate" phase. When the same load context is encountered again, the prefetcher assigns a walker to perform the walk once more to validate that the code slice did not change. The PIE remains in validation phase for several encounters to ensure the code is stable and to identify constant/strided values (the strides themselves may be caused by code beyond the scope of the slice).

The prefetcher performs three validation rounds. Other values (up to seven) were tested, indicating that prime numbers work better, especially when no BHR context is used, as they may avoid some loop patterns from confusing the validation process. However, a lower value was chosen as overall performance benefits more from the speed of generating new slices than from the additional accuracy that may accompany longer validation. 

After finishing all validation rounds the entry is switched to the "Trim" phase. The "Trim" phase is the only step allowed to change the PIE slice since it was first generated. It performs the same walk, but stops tracking sources when reaching constants or strides that were discovered during the validation passes and replaces them with a simple immediate move or add/sub. As a result, some branches of the data dependency flow may be removed from the PIE slice. 

Another change performed during trimming is renaming the destinations to temporary registers to avoid any side effects. The walker performs a forward traversal over the constructed slice and converts each destination register to the next available temp register. The conversions are recorded in the temporary register map. 
During the following traversal steps, all younger slice instruction will rename any matching sources to read from the corresponding temporary register. 
After trimming is done, the entry is switched to "Armed" state. 

We assume that the walker FSM can handle up to 8 instructions per cycle without exceeding timing restrictions (based on similar existing mechanisms like branch recovery walks), so the full history walk should take up to 16 cycles. However, to ensure feasibility and allow larger history queue sizes, our evaluation assumes that a walk may take up to 64 cycles. Since the prefetcher may encounter additional loads during that time, it may use several parallel walkers, assigned to generation or validation phases based on availability.

\begin{figure}[t]
	\centering
	\fbox{\includegraphics[width=1\columnwidth]{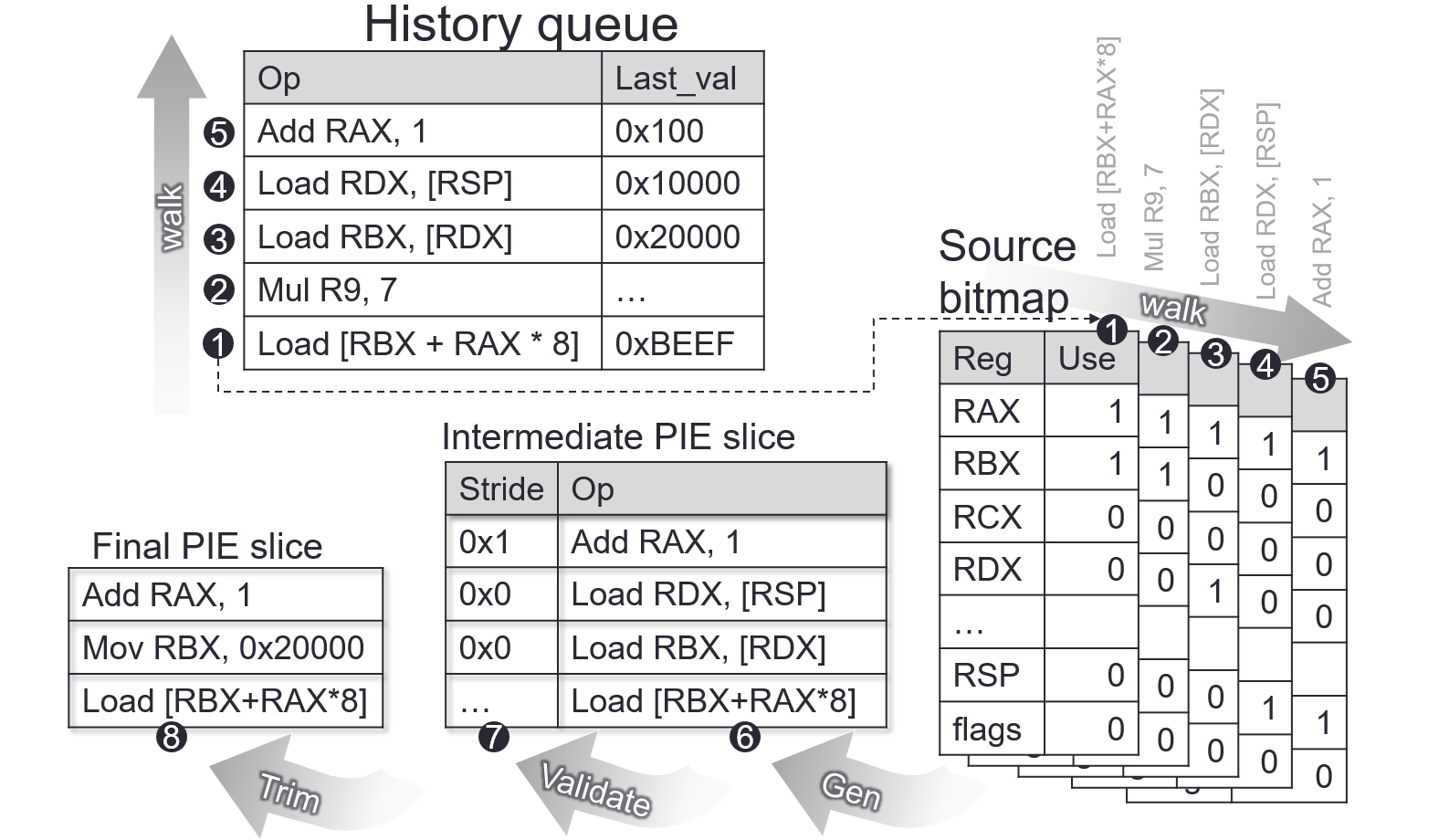}}
	\caption{Slice generation example. The history queue holds all instructions in the dynamic flow and is walked backwards from the triggering load. The sources bitmap is shown on the right during the walk (steps 1 through 5). After the walk we receive an intermediate PIE slice (6) including only the dependency chain (R9's multiply was dropped). We then populate the stride/const values during the validation steps (7). The final PIE slice shows the post-TRIM slice (8), in which RDX was discovered as constant and allowed eliminating the load fetching it.}
	\label{fig:slice_gen2}
\end{figure}

Figure~\ref{fig:slice_gen2} shows an example of slice generation over code striding across an array that requires double dereference (since, for e.g., it was passed by pointer). The dependency chain is discovered by walking the history queue as shown on the right hand side (removing the unrelated multiply operation but identifying all other operations as part of the dependency chain). The intermediate slice remains consistent during several validation phase iterations. During that phase the add operation is detected as a stride of one and the two middle loads are identified as constants.
Since RBX is constant, the Trim phase replaces it with a move and stops processing its dependencies (thereby also eliminating the RDX load). The final slice is therefore only three operations long.

\subsection{Slice injection}
Once a slice has been armed, each encounter with its load context (i.e., hitting the same IP while having the same branch history) triggers the PIE slice injection. The allocation stops immediately prior to the triggering load (thus preserving the same register roles and meaning as seen during construction). The slice operations are then injected in order, with no lingering side effects as the temporary registers used are not part of the architectural state. Any memory operation is allowed to lookup the TLBs and caches and, if needed, perform page walks and allocate line fill buffers. These accesses may, by themselves, act as prefetches. 

The injected operations may be executed by the normal machine out-of-order resources. 
However, this may incur a substantial cost to the actual program performance due to added stress over critical execution resources. Instead, an internal execution engine was added to perform the arithmetic operations without interfering with the normal core activity (other than stalling allocation). We evaluate both the shared-resources and the private-resources modes in Section~\ref{sec:evaluation}.
 
During the injection, sources marked as constants use the recorded constant value, but operations marked as having a stride are adjusted by having their stride value multiplied by a dynamic lookahead factor.
The dynamic lookahead is initialized for each slice to one, meaning that by default the prefetcher injects the PIE slice as-observed, with no extrapolation (thereby performing the address computation of the next iteration). However, if the PIE is eventually detected as non-timely (as explained in the next section) the lookahead factor will increase gradually up to a maximum of 64 (chosen to allow prefetching far enough ahead of time, but not too far as to exceed the cache lifetime). All strides within a slice are always multiplied by the same lookahead factor so that the ratios between strides are always kept as they were detected over a single iteration.

The final operation in the slice is a copy of the original load that the slice was constructed from, but since any strided sources were enhanced to apply a lookahead over their strides, the load address would belong to some future iteration. This becomes the final prefetch address and is sent to the memory unit as a prefetch. In parallel, it is also pushed to the \textit{prefetch queue} along with its predicting PIE-id for usefulness tracking.

\subsection{Usefulness tracking}
\label{sec:usefulness}
The generated prefetches must be correct and timely (i.e., the address should be used later by a demand, and do so within a sufficiently short time period as to avoid being flushed from the cache). We solve both requirements by tracking the prefetches in the prefetch queue. Each demand address is checked against the queue to find the first (most recent) matching prefetch, and the entry is marked as hit. If the hit is within useful distance (determined by a reward function as in the context-RL prefetcher~\cite{context_pref}), the PIE receives confidence upgrade based on the reward score. On the other hand, if a prefetch entry reaches the end of the prefetch queue without ever being hit, it is considered useless. 
The number of sent and useless prefetches is tracked in the PIE (the counters are both right-shifted whenever they are about to exceed in order to preserve their ratio). When a PIE goes below the usefulness threshold (we used 10\% in our experiments), it is reset, but allowed to regenerate the slice in case the current code flow changed compared to when it was originally constructed. If a PIE is reset more than 25 times, it is considered a stale PIE, and its state becomes \textit{Disabled}, preventing it from reconstructing.

Another form of filtering is tracking recurring addresses. The last address emitted by each slice is saved, and if the slice generates it again multiple times in a row, the slice is reset due to low usefulness.

\subsection{Dropping PIE slices}
Multiple issues could stop a slice construction process or reset an already constructed one. 
Construction can be aborted due to the following reasons:
	\begin{itemize}
	\itemsep0em 
	\item Slice is inconsistent during validation. 
	This may indicate insufficient context length, 
	the code having no useful recurrence, or a complex control flow. 
	\item Timeout while waiting for another validation pass, may indicate the load is not as hot as predicted. 
	\item Slice is too long (over 16 operations).
	\item Complex instruction (for e.g., non-trivial side effects). 
	\item Too many temporary registers (over 8) are needed. 
	\end{itemize}
Resets during slice construction are considered transient, meaning that a later attempt may still construct a useful slice. 
Conversely, slices can also be reset during run-time. 
	If too many prefetches fall off the prefetch queue without ever being hit by demands, the slice may have failed capturing the semantic relation. The minimal usefulness ratio is configurable and by default a threshold of 10\% is used. The failure rate is tracked using 2 counters: Failures and sent-prefetches. Both counters saturate at 64, and both shift right by 1 whenever any of them reaches that limit. If, after the counters reach steady-state, the ratio between them drops below the usefulness threshold, the prefetcher resets the entry.

Alternatively, If the same code slice produces the exact same address over and over, the slice is no longer considered meaningful. 
This may occur when reaching the history limit during construction,
when the walker cannot include a source being changed.
Aborting a slice at run-time provides information that triggering a prefetch on the initiating context might harm performance. 
Therefore, the PIE array records these resets and keeps them 
(unless the PIE is overridden by another load context). If too many run-time resets occur, the PIE switches to~\textit{disabled} state and no longer accepts re-construction attempts for that context.

\subsection{Prefetcher area}
The parameters used for the prefetcher are summarized in Table~\ref{table:pie_params}. For the sake of this paper each stored micro-operation is assumed to be represented with 64 bits including all data required for reproduction. The history queue entry also has to store the result 64-bit value (for const/stride detection), and therefore requires 128 bits. A 128-entries history queue requires 2kB.

Each PIE slice requires 16 operations, a context tag for indexing, a walker ID (used during generation and validation) and additional bits for state and reset tracking. Overall size is 140 Bytes. Since the PIEs are relatively large, the PIE array holds only 16 entries, with a total size of \tilde2.25kB. This is sufficient for most of the applications since the number of slices presented in Figure~\ref{fig:slice_count} refers to the entire lifetime of the application, but at any given program phase only a few slices are actively used. This is demonstrated later in Section~\ref{sec:resources}.
Future work may find ways to reduce the size of each entry (for example by compressing identical operations, as some slices may share parts of their history). The storage size is therefore not a fundamental issue. 

The slice generation FSM (walker) requires a
source bitmap (32 bits), a memory renaming cache (16 entries with a 64b tag + 4b index each = 136 bytes), and a temporary registers map (40 bits). Each FSM also has a slice storage for the construction process, reaching a total of \tilde280B. Having 2 parallel walkers would therefore require \tilde0.6kB.
 
Power consideration are reviewed in section~\ref{sec:evaluation}.

\begin{table}[t]
	\footnotesize
	\centering
	\begin{tabular}{| l | l |}
		\hline
		History queue & 128 instructions, 2kB \\
		\hline
		BHR size & 24 bit (4b $\times$ 6 last branches)\\
		\hline
		Mem. renaming cache & 16 $\times$ (64 + 4) bits = 1kB\\
		\hline
		Walkers  & 4 $\times$ 280B = 0.6kB\\
		\hline
		PIE array size & 16, 2.25kB\\    
		\hline
		Total size (kB) & 6kB \\    
		\hline
		hot/flaky thresholds &  2 appearances / 1 miss\\    
		\hline
	\end{tabular}
	\vspace{2mm}
	\caption{Prefetcher parameters}
	\label{table:pie_params}
\end{table}

\section{Methodology}
\label{sec:methodology}


\begin{table}[t]
	\scriptsize
	\centering
	\begin{tabular}{| p{50pt} | p{120pt} |}
		\hline
		Core type & OoO, 4-wide fetch, 3.2Ghz \\
		\hline
		Queue sizes & 224 ROB, 97 RS, 180/168 int/FP regs,\\
		& 72 load buffer, 56 store buffer \\
		\hline
		MSHRs &  (estimated)  10 L1, 20 L2\\
		\hline
		L1 cache & 32kB Data + 32kB Code, 8 way, 2 cycles\\
		\hline
		L2 cache & 256kB, 4 ways\\
		\hline
		L3 cache & 4MB, 16 ways x 2 slices\\
		\hline
		Memory & LPDDR3, 2-channel, 8GB, 1600 Mhz\\
		\hline
		\multicolumn{2}{c}{Competing prefetchers} \\
		\hline
		GHB (all)~\cite{GHB} & GHB size: 2K, History length: 3 \\
		  & Prefetch degree: 3, Overall size: 32kB \\
		\hline
		SMS~\cite{SMS} & PHT size: 2K, AGT size: 32, Filter: 32 \\
		& Regions size: 2kB, Overall size: 20kB \\
		\hline
		VLDP~\cite{VLDP} & 3 DPTs $\times$ 64 entries\\
		\hline
		Context RL~\cite{context_pref}  & CST size: 2K entries x 4 links (18kB), \\
		& Reducer: 16K entries (12kB) \\
		\hline
	\end{tabular}
	\vspace{2mm}
	\caption{Simulator parameters based on Skylake}
	\label{table:params}
\end{table}

The semantic prefetcher was implemented in a proprietary cycle-accurate x86 simulator configured to match the Skylake micro-architecture~\cite{skylake} and validated against real hardware over a wide range of applications (showing an average error margin within 2\%). Table~\ref{table:params} specifies the parameters used. All prefetchers support L1 triggering and virtual addresses and can use TLBs/page walks when needed.

The prefetcher was tested over the SPEC 2006 and 2017 benchmark suites, compiled on Linux with ICC 14 and 16 (respectively). Each application had 5-8 different traces chosen by a SimPoint-equivalent tool based on workload characterization. The traces are weighted to represent overall application performance while measuring 5M instructions each (following a warmup of about \tilde1B memory accesses and \tilde20-50M actual instructions). 


To test multithreaded performance we also run combinations of workloads on different threads, although we do not implement the ability to share the learnings across threads.
For that purpose, we use combinations of traces from the same applications to measure SPEC-rate behavior (where several copies of the same application are being run independently). We run the traces with separate physical memory address ranges to avoid data collisions. We also offset the run phases by a few million instructions to ensure some heterogeneity. 


\section{Evaluation}
\label{sec:evaluation}

\begin{figure}[t]
	\includegraphics[width=1.03\linewidth]{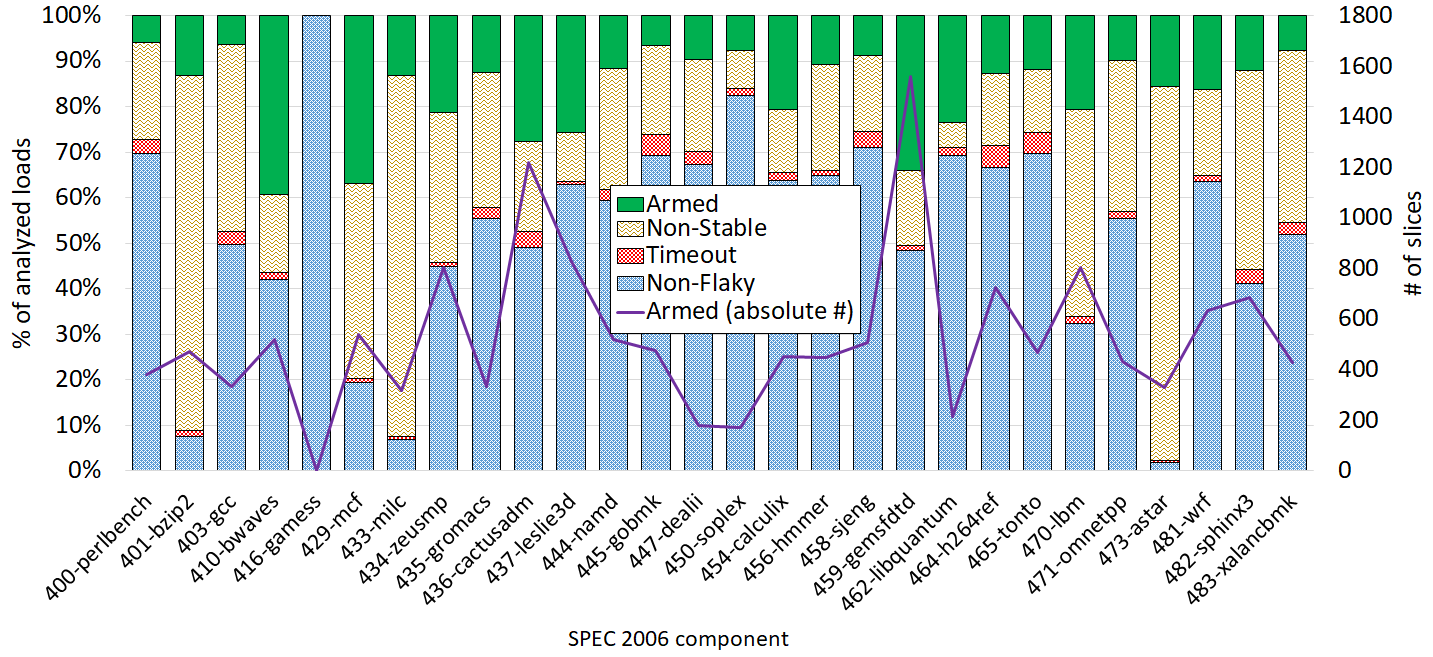}
	\\~
	\includegraphics[width=1.03\linewidth]{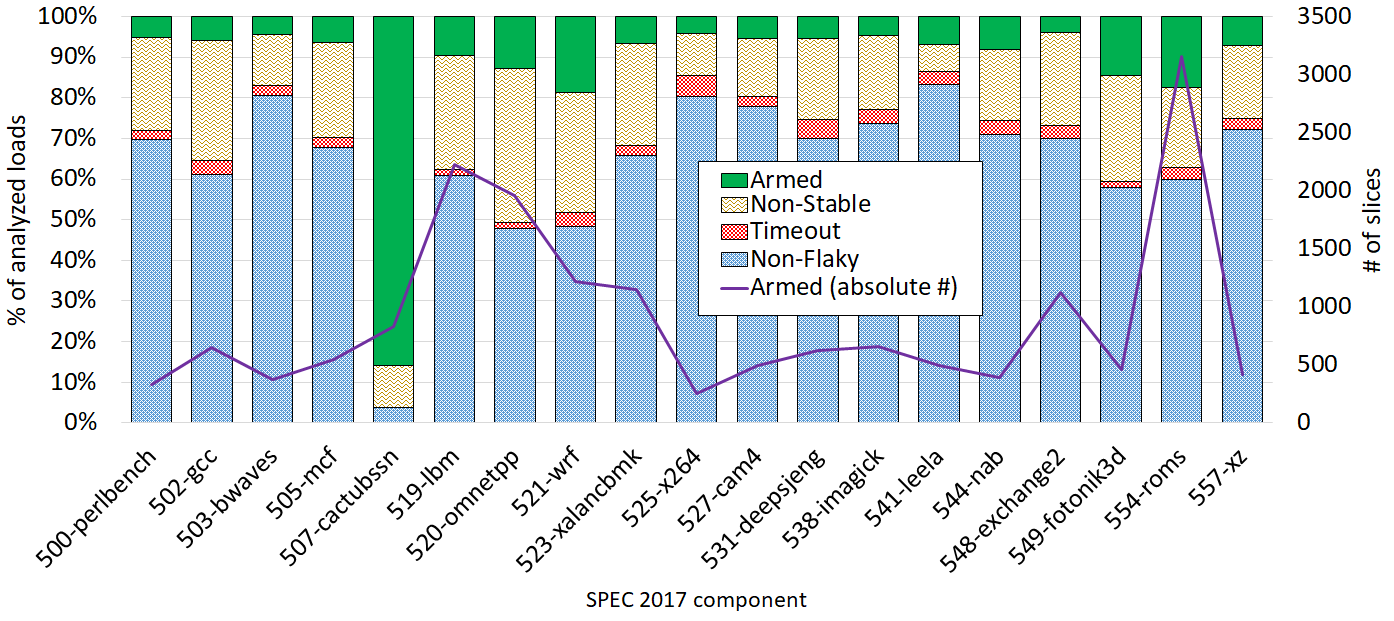}
	\caption{Slice coverage over SPEC-2006/2017. The left Y-axis measures the normalized portion of each outcome of slice construction. The right Y-axis (and overlaid line) show the absolute count of successfully armed slices.}
	\label{fig:coverage}
\end{figure}

\begin{figure}[t]
	\centering
	\includegraphics[width=0.8\linewidth]{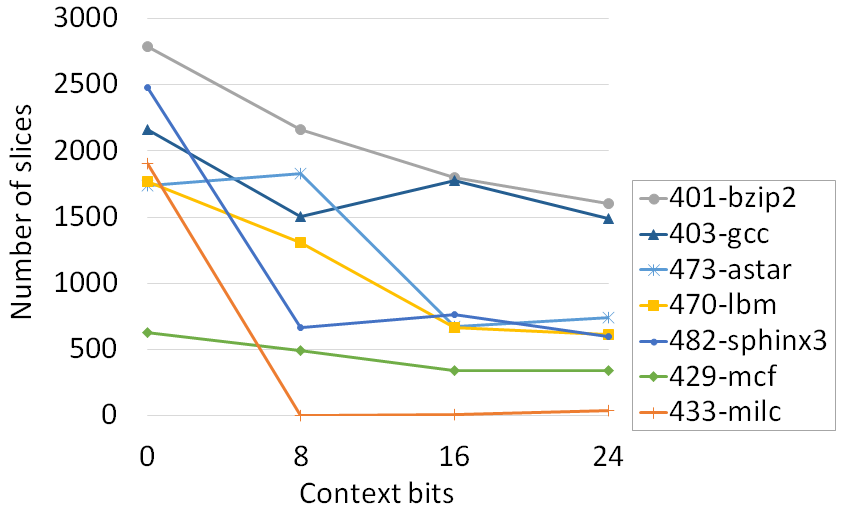}
	\caption{Number of slices failing validation per each context length. Longer context contributes to the consistency of slices (since iso-context slices are more likely to share the same locality behavior) }
	\label{fig:context}
\end{figure}

\begin{figure}[ht]
	\hspace{-5mm}
	\includegraphics[width=1.05\linewidth]{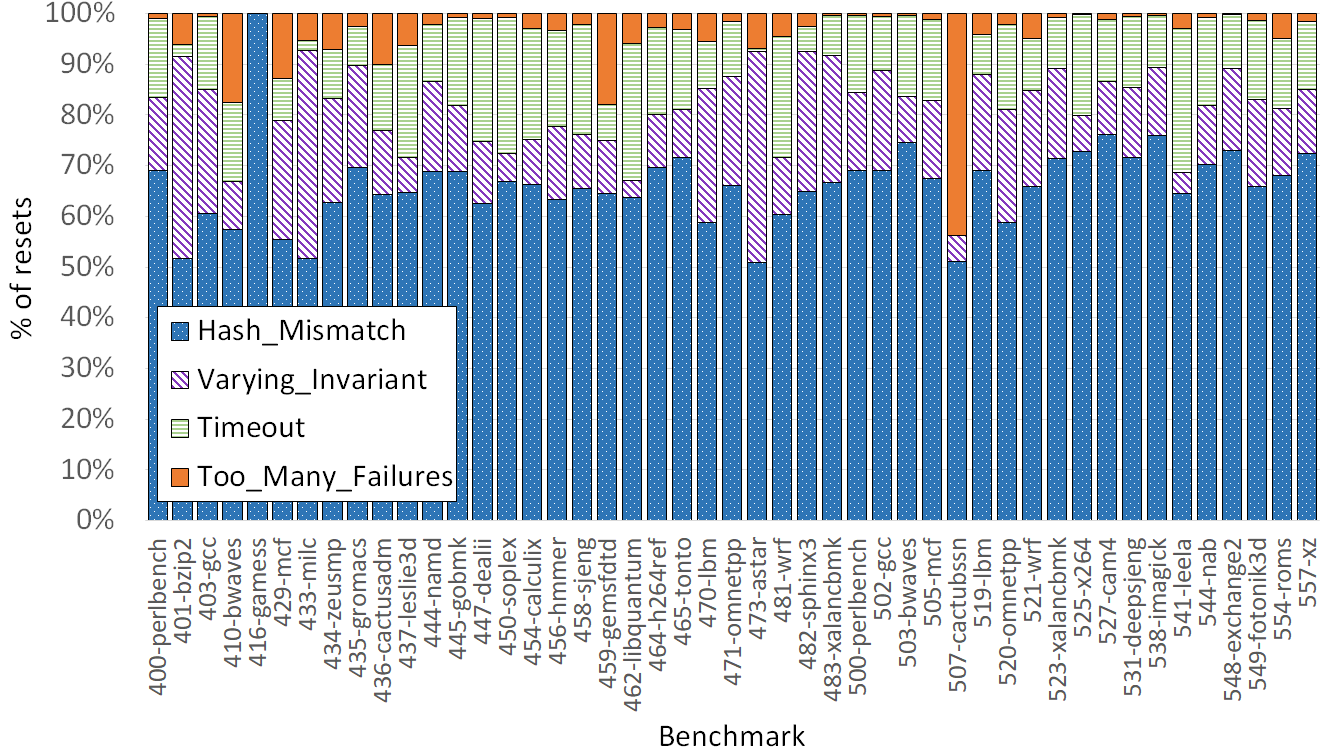}
	\caption{Breakdown of reset causes.}
	\label{fig:resets}
\end{figure}

\begin{figure}[ht]
	\hspace{-5mm}
	\includegraphics[width=1.03\linewidth]{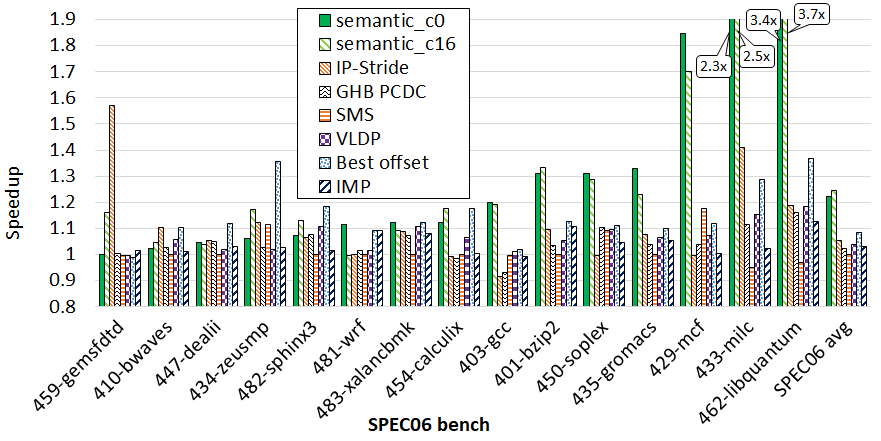}
	\hspace{-5mm}
	\includegraphics[width=1.0\linewidth]{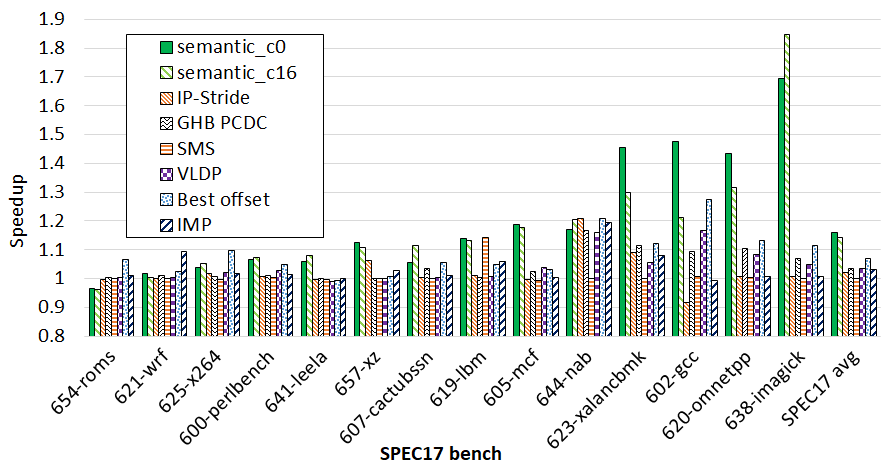}
	\caption{Semantic prefetcher speedup vs competition. Showing workloads with any prefetcher gaining over 7\% (average is over full SPEC Int/FP).}
	\label{fig:speedup}
	\vspace{-2mm}
\end{figure}

To evaluate the benefits of the semantic prefetcher, we first need to determine its ability to cover enough performance-critical loads within common workloads. Figure~\ref{fig:coverage} shows the coverage of different SPEC 2006/2017 workloads: each application shows the number of dynamic loads analyzed by the semantic prefetcher, and the break-down by analysis outcome. The \textit{Non-Flaky} component counts loads not deemed hot enough or not having enough misses to be interesting. The \textit{Timing} component shows slices failing during the validation period due to low rate of recurrence or hash conflicts. 
The \textit{Non-Stable} component shows slices failing validation due to variability of code path (results are shown with zero context, increasing the context is shown later to reduce this component as the code paths are more consistent when compared across recurrences with the same branch history). 
The remaining component at the top shows the number of armed slices per workload. 

The absolute number of slice validation failures is shown in Figure~\ref{fig:context} for the 7 SPEC workloads that have the highest failure rate when using no context (all having over 40\% of their slices reset during slice generation). The number of failures is compared having between 0 to 24 context bits (i.e., indexing loads based on the history of the last 0 to 6 branches and their resolutions). Adding the full context length reduced between 30\% (gcc) to 98\% (milc) of the failures, indicating that recurrences with the same branch history are more likely to have consistent code slice behavior.

The overall breakdown of reset causes appears in Figure~\ref{fig:resets}. The first element is failures due to hash collisions: new dynamic loads matching the PIE index of an existing slice that is under construction, causing it to drop (armed slices are protected from overwrite and can only be reset due to low usefulness). The second element is variance in code flow during slice generation. The third is timeout during the construction: a slice that was not armed within 100k cycles is reset due to low relevance. 
The last reset cause, Too-Many-Failures, is a run-time reset cause, occurring after a slice was validated and armed, as explained in Section~\ref{sec:usefulness}. 

It should be noted that the various reset thresholds and parameters have shown very little sensitivity to tuning. This happens because most slice stabilization and resets occur during warmup and are therefore negligible on longer runs. 
On the other hand, flakiness and usefulness parameters are more sensitive to tuning, and show better performance the more aggressive they are dialed (i.e., building and maintaining slices for more loads). However, optimizing with a higher cost of slice injection (especially without dedicated execution) would likely lead to more conservative thresholds. 

The speedup of the semantic prefetcher (with 0 and 16 bits context) is shown in Figure~\ref{fig:speedup} across SPEC 2006/2017 benchmarks. Several competing prefetchers are also shown. 
The overall gain is significantly higher on SPEC 2006 (24.5\%), mostly due to the lack of software prefetching by the older compiler, but the improvement exists also on SPEC 2017. 

\begin{figure}[t]
	\includegraphics[width=1\linewidth]{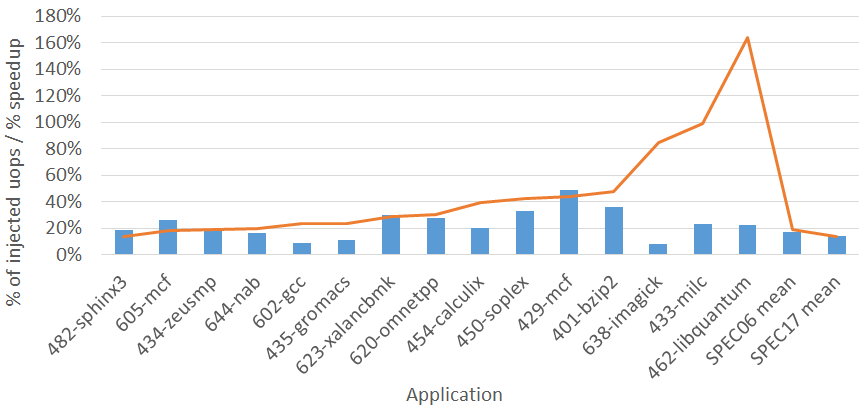}
	\caption{Ratio of injected operations out of the overall, compared to speedup. Showing applications with speedup $\geq$ 15\%.}
	\label{fig:uops_added}
\end{figure}

Slice injection also adds computation work. Figure~\ref{fig:uops_added} shows the injected instructions out of the overall instructions executed, compared with the performance gain. In most cases there is good correlation (i.e., the performance gain is proportional to the added work), but some applications with relatively simple slices are able to gain significantly more than their overhead. If we assume the prefetcher's steady-state power cost (disregarding slice generation) is equivalent to the added operations, then the power/performance score has on average 2.5$\times$ more IPC gain than power cost.

\begin{figure}[]
	\includegraphics[width=0.9\linewidth]{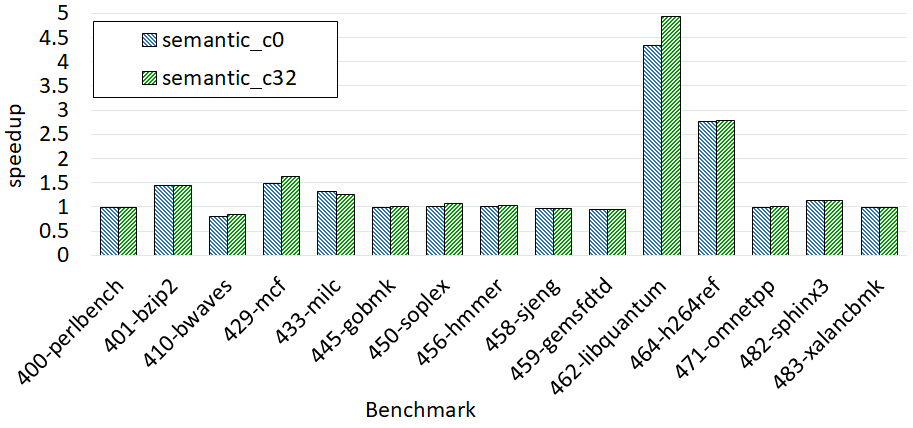}
	\caption{Semantic prefetcher speedup over MT SPEC06 workload combinations (4 cores).}
	\label{fig:mt_speedup}
	\vspace{-3mm}
\end{figure}

Finally, the semantic prefetcher improves performance also on multi-threaded runs. Figure~\ref{fig:mt_speedup} shows the speedup over SPEC-rate simulation (4 traces from the same application over 4 physical cores). In some cases prefetching provides a higher gain than on a single thread (in h264, for example). 
There is no sharing of generated slices across physical cores, so the only gain comes from increasing the effective prefetching depth. On MT runs the system becomes more saturated and memory bandwidth usually becomes a more critical bottleneck compared to memory latency. This may reduce the efficiency of prefetching (or even the chances of issuing the prefetches). On the other hand, prefetches may also serve as cache hints that increase the lifetime of required lines, thereby reducing overall bandwidth and improving MT performance. On the highest MT gainer (libquantum), the prefetcher reduced L1 MPKI by 40\%. 

\subsection{Comparison with other prefetchers}
We compare the speedup of the semantic prefetcher with other prefetcher with different approaches and coverage. The semantic prefetcher wins over most SPEC workloads, scoring on average more than twice the speedup of the next best prefetcher. However, on some workloads the semantic prefetcher loses to one of the competing prefetchers. 
In gemsFDTD, a simple stride prefetcer is able to gain almost 60\% speedup while the semantic prefetcher gains only \tilde17\% at best. The reason for that is short nested loops, where the inner recurrence is too long to fit in the context history length, but too short to allow the chance for re-learning the inner loop on every outer iteration. This control flow gives an advantage to simple and fast prefetchers that need only a few loop recurrences to learn and trigger (the stride prefetcher can start issuing prefetches after the 3rd inner iteration). The semantic prefetcher can still learn the code pattern given sufficient context, and in fact begins to gain performance by covering at least some of the cases with a context of 32 bits and above, but such context length begins to stress the physical design and does not solve the general case where inner loops can be much longer. This can be solved by having the context support compressed representation of loop patterns. 
 
Another competing prefetcher that gains over the semantic prefetcher over some workloads is the Best-Offset prefetcher. BOP has several outliers in its favor, most notably zeusMP. Unlike GHB and other fast-learning prefetches, BOP also takes a while to discover the optimal depth, but once it does, it has a throttling effect where it eliminates unnecessary prefetches (at sub-optimal offsets). The gains in zeusMP (and to a lesser extent also in dealII, sphinx3 and cactusADM) are mostly through reduction of BW from excessive prefetching. For the same reason adding context to the semantic prefetcher also helps on zeusMP by eliminating badly constructed slices emitting some useless prefetches.

Finally, IMP presents an interesting point: within the SPEC applications it wins only on WRF (and only by 8\%), but the graph500 example had array dereferences that make IMP quite useful, except on the longest ones. 

\subsection{Lookahead method}
\begin{figure}[]
	\includegraphics[width=0.95\linewidth]{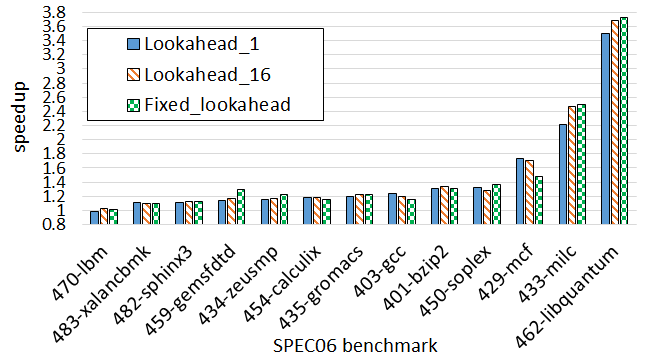}
	\vspace{-2mm}
	\caption{Gain with different lookahead policies.}
	\label{fig:lookaheads}
\end{figure}

The lookahead multiplier (the number of strides we prefetch ahead) plays a key role in the prefetcher speedup. However, when applying a fixed multipliers we noticed that different workloads were favoring different values, and optimizing the best method required dynamic tuning. We implemented the following approaches:
\begin{itemize}
	\item Constant lookahead: we set a constant value and always perform the lookahead according to it.
	\item Hit-depth normalization: we measure the average depth of hits within the prefetch queue which indicate the actual distance between a prefetch and the demands access using it. We then increase or decrease the lookahead value to normalize this hit depth to the desired range (if we hit too early, around the beginning of the prefetch queue, we need to extend our lookahead and vice versa).
\end{itemize}
The difference between the policies is shown in Figure~\ref{fig:lookaheads}. Lookahead 1 and 16 are dynamically adjusting the lookahead distance while starting from a multiplier of 1 or 16 iterations ahead (respectively) and increasing from there. The fixed lookahead policy (always 32 iterations ahead) has some minor gains in cases where it starts from a more effective prefetching depth while the dynamic policy takes a while to reach there, but it is ultimately inferior on most runs where the dynamic approach is more adaptable.

\subsection{Execution resources}
\label{sec:resources}
\begin{figure}[]
	\centering
		\includegraphics[width=1\linewidth]{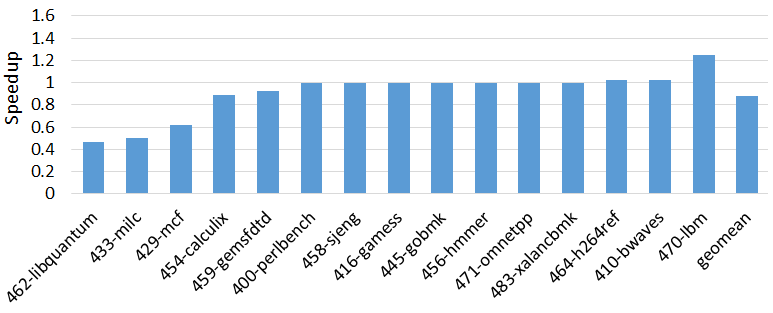} %
		\vspace{-6mm}
		\caption{Using existing vs. dedicated ALU.}
		\label{fig:core_resources}
\end{figure}

The semantic prefetcher uses a dedicated generic ALU as the execution engine of the PIE slices. However, if the slice execution latency becomes a critical factor, the area addition is too high, and the overall execution bandwidth is sufficient, we may choose to simply inject the slice code into the main out-of-order engine and let the existing resources do the work for us. Figure~\ref{fig:core_resources} shows the penalty of dropping the dedicated HW and sharing the core execution resources.

\begin{figure}[]
	\centering
		\includegraphics[width=0.8\linewidth]{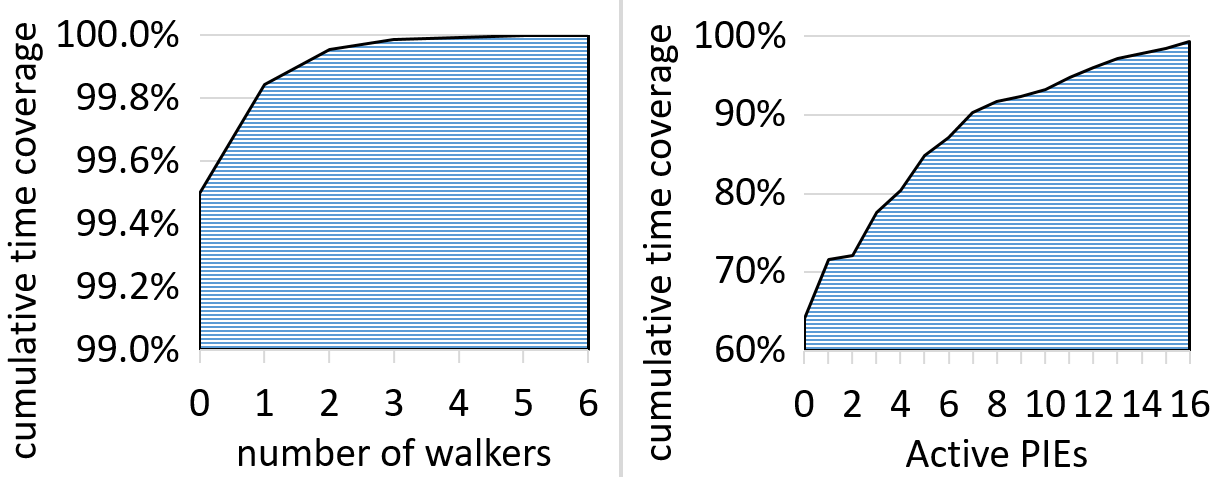}
		\caption{Amount of parallel PIEs and active walks required per cycle.}
		\label{fig:num_walkers}
\end{figure}

Another tradeoff is the number of walkers performing the slice generation and validation. Figure~\ref{fig:num_walkers} shows how many parallel walks (traversals over the history queue) and PIE entries (slices tracked for constructions) are needed using cumulative time histograms. Overall cycles with any number of walks consume only 0.5\% of the run-time, (also indicating that the power consumption of the walk itself is negligible). The results indicate 
that two walkers are sufficient. In the same way, 16 PIE entries are enough to cover 99.4\% of the run. 

\vspace{-2mm}
\section{Related work}
\label{sec:related}

\subsection{Program semantics}
Multiple researches attempt to automate the analysis and understanding of software applications. \textit{Shape analysis}~\cite{shape_analysis} attempts to build a set of properties and rules representing the program's data sets in order to facilitate  program verification (mostly of memory object allocation and management, bounds/coherence checking and functional correctness). 

Newer approaches attempt to represent programs in abstract forms derived from their code and behavioral analysis~\cite{code_vectors,inst2vec}, in order to find similarities for code suggestion/completion, anti-plagiarism, or algorithm identification.
These approaches may be useful in high level dynamic tuning (adjusting HW properties such as the type of prefetcher used, or optimal aggressiveness), but they do not yet assist in the analysis of the access pattern or address generation.

\subsection{Using code slices}
Collecting and using slices of actual code has already been proposed for various purposes. Trace cache~\cite{trace_cache} is a form of constructing and efficiently caching selective code paths based on run-time analysis.
In the realm of branch prediction, Zilles et al.~\cite{slice_prediction} proposed using code slices to execute ahead the predicted code path to resolve hard-to-predict branches. A similar approach suggested by Peled et al.~\cite{FBPQ} is based on injecting code slices to resolve data-dependent branches by prefetching the data, using it to resolve the branch condition, and queuing the resolution for overriding the prediction.

This method was also proposed for memory latency mitigation. Carlson et al.~\cite{load_slice} proposed a similar mechanism that dynamically learns load dependency chains in order to expedite their execution. However, their approach was based on in-order cores, and motivated to extract a small portion of the ILP available to full out-of-order cores by execution only load address dependencies out of order. 

Prefetching can also be achieved by executing actual code (or even just critical subsets of it) ahead of time as proposed by Mutlu and Hashemi et al. in their set of Runahead techniques~\cite{runahead_ooo,continuous_runahead, filtered_runahead}, and by Collins et al. in their speculative precomputation technique~\cite{precompute} (extended by Atta et al.~\cite{precompute2} to include also limited control flow). That work relied on continued execution of the same program context past memory stalls, and focused on managing a complicated speculative execution state for that purpose. It did not modify the executed code but most approaches did filter out code not required for memory address calculation. It also did no extrapolation, and thus was limited in range to what could fit in the enhanced out-of-order window.

Prefetching based on actual code slices can also be done by helper threads (Runahead flavor by Xekalakis et al.~\cite{runahead_smt} and slice-based processors by Moshovos et al.~\cite{slice_proc}). Similar decoupled approaches were also used for branch prediction by Chappell et al.~\cite{SSMT, SSMT2} and Farcy et al.~\cite{farcy1998dataflow}. However, this form is asynchronous with the progress of the main thread and will not be able to ensure fixed (or even positive) prefetching distance.

Another form of expedited execution through other threads is ~\textit{Hardware scouting}~\cite{scouting}, which is intended for highly multithreaded machines and uses other threads to run ahead of execution. However this approach attempts to optimize MT throughput, and not address single-threaded performance.

\subsection{Prefetching techniques}
Falsafi and Wenisch classified prefetching techniques into the following groups~\cite{PrimerHWPref}:
\begin{itemize}
\itemsep0em
\item \textbf{Stream/stride prefetchers} utilize \textit{spatial locality}, by detecting constant stride patterns. The sandbox prefetcher~\cite{sandbox}, 
the best-offset prefetcher (BOP)~\cite{BOP}, 
and Access Map Pattern Matching (AMPM)~\cite{AMPM}, 
proposed various methods of testing different strides and choosing the optimal one, thereby covering complex flows through common recurrence deltas. Other prefetchers such as the variable length delta prefetcher (VLDP)~\cite{VLDP} 
enhanced that ability to varying stride patterns.

\item \textbf{Address-correlating prefetchers} detect correlation within sequences of recurring accesses. 
This form of locality has the ability to cover some semantic relations, but is ultimately limited to the storage capacity of correlated addresses. 
Examples include the Markov predictor~\cite{MarkovPredictors}, the Global History Buffer Address-Correlation flavors (GHB/AC) \authorhide{by Nesbit and Smith}~\cite{GHB}, 
and prefetchers targeting linked data structures through partial memoization, such as that by Roth, Moshovos and Sohi~\cite{Roth98,Roth99}, and Bekerman et al.~\cite{Bekerman99}. 
An extension of address-correlation is context-correlation 
~\cite{context_pref,nnpref} which seeks to correlate a larger context vector with future addresses. 

\item \textbf{Spatially-correlated prefetchers} use an extension of  temporal locality that correlates between spatial patterns instead of absolute addresses. These prefetchers seek out recurring spatial patterns that are not part of a long consecutive sequence but may repeat locally, such as accesses to the same fields of a structure across different instances. Examples of this family are Spatial Memory Streaming (SMS)\authorhide{by Somogyi et al.}~\cite{SMS} and the DC flavors of GHB~\cite{GHB}.

\item \textbf{Irregular data patterns prefechers} target specific data structured that do not have spatio-temporal locality. IMP\authorhide{by Yu et al.}~\cite{IMP} prefetches future elements within an array of indexes (A[B[i]]). 
Other data-driven prefetchers include the Irregular Stream Buffer (ISB)\authorhide{by Jain and Lin}~\cite{ISB}, which restructures the dataset spatially. 
Another form of irregular prefetching based on context if B-Fetch~\cite{BFetch} by Kadjo et al. which uses branch history to detect strides in registers used for address generation. 
However, since it does not execute actual code it is limited to simple register strides and cannot reconstruct complex value  manipulations or see through memory dereferences.
\end{itemize}



\section{Conclusions}
\label{sec:conclusions}

This paper presents the semantic prefetcher. The prefetcer is designed to utilize the most generalized form of locality: \textit{semantic locality}, which is not limited to spatial or temporal artifacts of the memory address sequence, but instead attempts to extract the code slices responsible for invoking consequential memory accesses along the data structure traversal path or algorithmic flow. We then combine and manipulate these slices into forecast slices by utilizing locality artifacts within their code to create new functionality that can predict the memory behavior of future iterations.

While some existing prefetchers attempt to capture access patterns that belong to specific use-cases, whether spatio-temporal relations, temporal correlations, or even irregular and data-dependent patterns, there is currently no generalized technique that can capture all such cases. 
The semantic prefetcher attempts to solve that by observing the code path directly, imitating any program functionality and extending it to create lookahead functionality. 
Based on that technique, the semantic prefetcher can extend the coverage provided by existing prefetchers (including irregular ones) through inherently supporting complex address generation flows and multiple memory dereferences.
The semantic prefetcher provides a speedup of 24.5\% over SPEC-2006 and 16\% over SPEC-2017, exceeding gains from other prefetchers by over 2$\times$. 




\bibliographystyle{IEEEtranS}
\bibliography{refs}

\end{document}